# Carrier loss mechanisms in textured crystalline Si-based solar cells

Akihiro Nakane, Shohei Fujimoto, and Hiroyuki Fujiwara[a]

Department of Electrical, Electronic and Computer Engineering, Gifu University, 1-1 Yanagido, Gifu 501-1193, Japan

**Abstract**

A quite general device analysis method that allows the direct evaluation of optical and recombination losses in crystalline silicon (c-Si)-based solar cells has been developed. By applying this technique, the optical and physical limiting factors of the state-of-the-art solar cells with ~20% efficiencies have been revealed. In the established method, the carrier loss mechanisms are characterized from the external quantum efficiency (EQE) analysis with very low computational cost. In particular, the EQE analyses of textured c-Si solar cells are implemented by employing the experimental reflectance spectra obtained directly from the actual devices while using flat optical models without any fitting parameters. We find that the developed method provides almost perfect fitting to EQE spectra reported for various textured c-Si solar cells, including c-Si heterojunction solar cells, a dopant-free c-Si solar cell with a $MoO_x$ layer, and an n-type passivated emitter with rear locally diffused (PERL) solar cell. The modeling of the recombination loss further allows the extraction of the minority carrier diffusion length and surface recombination velocity from the EQE analysis. Based on the EQE analysis results, the carrier loss mechanisms in different types of c-Si solar cells are discussed.

## I. INTRODUCTION

Pyramid-shaped textures with sizes of 5~10 μm are generally incorporated into crystalline silicon (c-Si) solar cells to suppress the front light reflection and thus to enhance the light absorption in the devices.[1,2] The large textures formed on the front and rear surfaces of the c-Si, however, complicate the optical analysis significantly, making the determination of the carrier loss mechanisms within the devices quite challenging.

The difficulty of performing the explicit optical characterization arises particularly from the randomness of the pyramid textures, formed generally by alkaline wet etching of c-Si (100) wafers.[1-3] So far, to characterize the light trapping properties of various c-Si textures, a computer-intensive ray tracing technique has been applied,[1,4-11] but these studies provide limited success, as the full optical analysis of the multilayered c-Si device has been rather difficult due to the large calculation cost of this approach. Thus, there is still a strong need for the development of a novel optical simulator that can be employed for the practical characterization of textured c-Si devices on a routine basis.

Such a technique is critical for the efficient optimization of solar cells. In particular, in c-Si heterojunction solar cells, hydrogenated amorphous silicon (a-Si:H) and transparent conductive oxide (TCO) layers exhibit large unfavorable parasitic absorption, reducing the external quantum efficiency (EQE) in the short and long wavelength ($\lambda$) regions notably.[9-13] For the heterojunction solar cells formed on flat c-Si substrates, detailed EQE characterization has been performed to determine the optical losses in the component layers;[11] however, for c-Si solar cells with random textures, the complete optical-loss analysis has not been reported yet.

On the other hand, the effect of the carrier recombination appears clearly in the EQE spectra of c-Si solar cells, and the intensive carrier recombination observed at a c-Si/Al rear interface reduces the EQE response in the longer $\lambda$ region remarkably.[14] Accordingly, all the optical and recombination losses in the solar cells can be assessed quantitatively based on the EQE analysis if the proper analysis method is established.

In our previous study,[15-17] we have established an EQE analysis technique for thin-film solar cells with submicron textures. In this method, to determine the light absorption in solar cells accurately, reflectance spectra obtained experimentally have been applied assuming flat optical models within the framework of the optical admittance method.[18] This method provides excellent fittings to numerous EQE spectra reported for Cu(In,Ga)Se$_2$ (Ref. 15), Cu$_2$ZnSn(S,Se)$_4$ (Ref. 17), and hybrid perovskite[16,19] solar cells, enabling the accurate characterization of the carrier loss mechanisms in these devices.

In this study, to reveal the optical and physical limiting factors of various c-Si-based solar cells, we have developed a global EQE analysis method in which the light absorption in the c-Si with a random texture is assessed using the experimental reflectance spectrum while assuming a perfectly flat optical model. By this procedure, the EQE calculations of the textured structures are simplified drastically. To reproduce the incoherent light absorption observed in thick c-Si wafers, a calculation scheme has also been established. As characterization examples, we present the EQE analyses for (i) c-Si heterojunction solar cells,[9,14,20] (ii) a dopant-free c-Si solar cell[21] and (iii) an n-type passivated emitter with rear locally diffused (PERL) solar cell.[22] We have further developed recombination analysis that can be incorporated into the above EQE analysis. From our approach, all the reflection, parasitic absorption and recombination losses in c-Si solar cells can be evaluated systematically.

## II. EXPERIMENT

In this study, the EQE spectra of high-efficiency c-Si solar cells reported earlier[9,14,20-22] have been analyzed. Figure 1 shows the structures of the c-Si solar cells analyzed by the developed EQE analysis method: flat c-Si heterojunction solar cells with (a) single-hetero (SH) and (b) double-hetero (DH) structures,[14] (c) a textured a-Si:H/c-Si heterojunction solar cell,[20] (d) a textured dopant-free $MoO_x$/c-Si solar cell,[21] and (e) a textured n-type PERL solar cell.[22] In this figure, the layer thicknesses of the solar-cell component layers used in the EQE analyses are also indicated. These values were adopted from the descriptions in the references. In the analyses of Fig. 1(c) and 1(d), however, the a-Si:H layer thicknesses were adjusted slightly to obtain the better matching with the experimental results, and the original thicknesses are shown inside the parentheses in Fig. 1. In Table I, the short-circuit current density ($J_{sc}$), open-circuit voltage ($V_{oc}$), fill factor (*FF*) and conversion efficiency of the above solar cells, determined from the current density-voltage characteristics, are summarized.

In the solar cells of Figs. 1(a) and 1(b), hydrogenated amorphous silicon oxide (a-SiO:H) layers are introduced, instead of conventional a-Si:H layers, to suppress the detrimental epitaxial growth of an intended a-Si:H layer on c-Si.[23] The elimination of the epitaxial phase at the a-Si:H/c-Si interface is critical to maintain high $V_{oc}$ (Ref. 24) and the epitaxial phase formation is suppressed quite effectively when a-SiO:H is employed.[23] For the solar cells of Figs. 1(a) and 1(b), the O contents of 7 at.% (p and n layers) and 4 at.% (i layer) were employed.[14] The conversion efficiencies of

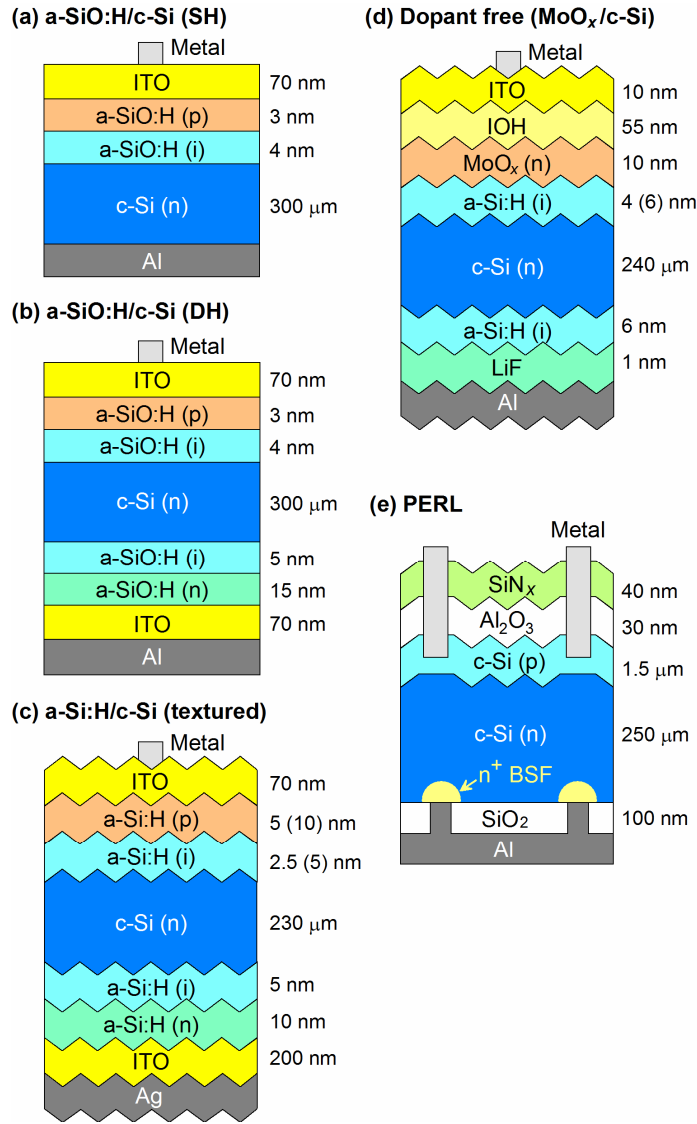

FIG. 1. Structures of c-Si solar cells analyzed in this study: flat c-Si heterojunction solar cells with (a) single-hetero (SH) and (b) double-hetero (DH) structures,[14] (c) a textured a-Si:H/c-Si heterojunction solar cell,[20] (d) a textured dopant-free MoO$_x$/c-Si solar cell,[21] and (e) a textured n-type PERL solar cell.[22] The layer thicknesses of the solar-cell component layers adopted from the references are indicated and these thicknesses were used in the actual EQE analyses. For (c) and (d), however, the a-Si:H layer thicknesses were adjusted slightly and the original thicknesses are shown inside the parentheses.

TABLE I. Characteristics of c-Si solar cells shown in Fig. 1.

| Device | Structure | $J_{sc}$ (mA/cm$^2$) | $V_{oc}$ (mV) | FF | Efficiency (%) |
|---|---|---|---|---|---|
| a-SiO:H/c-Si (SH) [a] | Fig. 1(a) | 33.1 | 643 | 0.76 | 16.2 |
| a-SiO:H/c-Si (DH) [a] | Fig. 1(b) | 35.6 | 656 | 0.75 | 17.5 |
| a-Si:H/c-Si [b] | Fig. 1(c) | 37.0 | 718 | 0.78 | 20.7 |
| MoO$_x$/c-Si [c] | Fig. 1(d) | 37.1 | 716 | 0.73 | 19.4 |
| PERL [d] | Fig. 1(e) | 41.2 | 703 | 0.80 | 23.2 |

a) Ref. 14, b) Obtained from Fig. 7 of Ref. 20, c), Ref. 21, d) Ref. 22

a-SiO:H/c-Si solar cells are comparable to those of the a-Si:H/c-Si solar cells.[20,23] In Table I, $J_{sc}$ of the SH a-SiO:H/c-Si solar cell is lower, compared with the DH structure, due to the intense carrier recombination at the c-Si/Al rear interface. As known well,[1] this rear carrier recombination can be reduced drastically by the introduction of a back surface field (BSF) structure [i.e., a-SiO:H(i)/a-SiO:H(n) layers in Fig. 1(b)].

The textured a-Si:H/c-Si solar cell in Fig. 1(c) represents a standard structure of c-Si heterojunction solar cells. Due to the presence of a pyramid-type texture, $J_{sc}$ of this solar cell is higher than that of the flat-type solar cell by ~2 mA/cm$^2$. The dopant-free solar cell architecture [Fig. 1(d)] has been proposed to eliminated the $J_{sc}$ loss caused by the parasitic absorption of the a-Si:H p layer in Fig. 1(c), although the a-Si:H i layer is still necessary to maintain high $V_{oc}$.[21] In this advanced structure, a high work function MoO$_x$ layer is employed as a front contact layer, whereas a LiF tunneling layer is provided as a rear contact layer. Moreover, to suppress the free carrier absorption in TCO layers,[25] the front TCO of this solar cell has a bilayer structure[26] consisting of a high-mobility In$_2$O$_3$:H (IOH) layer[27,28] and a conventional In$_2$O$_3$:Sn (ITO) layer. It should be noted that the ITO layer in this device is a contact layer, which is employed to reduce contact resistance with the metal electrode.[26]

The PERL solar cell [Fig. 1(e)] was fabricated using a surface texture with inverted pyramids.[22] In this device, SiN$_x$ antireflection and Al$_2$O$_3$ passivation layers are formed on the front surface of the p-type emitter (140 Ω/sq). The rear surface of this solar cell is passivated by SiO$_2$ and the rear electrical contact is made through the contact hole where n$^+$ diffused region is created as the local BSF structure. This solar cell shows a quite high conversion efficiency of 23.2% with higher $J_{sc}$ than those of the heterojunction cells due to the absence of the TCO and a-Si:H-based layers. However, $V_{oc}$ of the PERL cell is smaller, particularly when compared with a record-efficiency a-Si:H/c-Si solar cell ($V_{oc}$=743 mV)[29], as the heterointerface suppresses the interface recombination and the saturation current density effectively.[12,13]

## III. GLOBAL EQE ANALYSIS METHOD

### A. Calculation of flat solar cells

The EQE calculation in this study is based on the optical admittance method,[18] in which a flat optical model is assumed. Figure 2 shows the flat optical model and the calculation procedure of the EQE spectra. In the optical model, the $j$th layer is assumed to be an optically thick incoherent layer. In the figure, $N$ represents the complex refractive index ($N = n - ik$) defined by the refractive index $n$ and the extinction coefficient $k$.[30] The optical admittance $Y$ is expressed as $Y = H_f/E_f$, where $H_f$ and $E_f$ show the magnetic and electric fields, respectively. As known well,[18,30] there is a relation of $H_f = NE_f$ and thus $Y$ basically corresponds to $N$. In the case of Fig. 2, we obtain $Y_j = N_{j+1}$.

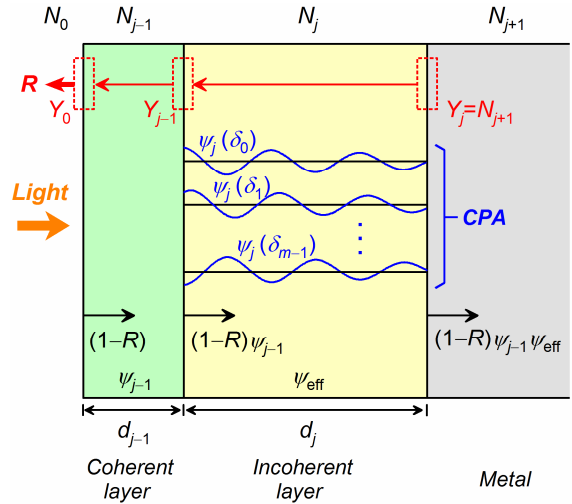

FIG. 2. Calculation procedure of the CPA method. In this method, $\psi$ of the incoherent layer ($\psi_j$) is calculated for various $\delta_p$ given by Eq. (6) and the effective $\psi$ ($\psi_{eff}$) is obtained as an average of $\psi_j(\delta_p)$. The slight attenuation of the wave amplitude indicates the light absorption in the incoherent layer.

In the conventional optical admittance method applied for optically coherent systems, $Y_j$ is transferred to $Y_{j-1}$ according to

$$Y_{j-1} = \frac{Y_j \cos\delta_j + iN_j \sin\delta_j}{\cos\delta_j + iY_j \sin\delta_j / N_j}, \qquad (1)$$

where $\delta_j$ shows the phase thickness expressed by

$$\delta_j = 2\pi N_j d_j / \lambda. \qquad (2)$$

Here, $d$ and $\lambda$ are the layer thickness and wavelength, respectively. If the calculation of Eq. (1) is repeated in a multilayer structure, we obtain $Y_0$ from which the reflectance ($R$) of the optical model is calculated as

$$R = |1 - Y_0|^2 / |1 + Y_0|^2. \qquad (3)$$

The important feature of the optical admittance method is that the transmittance ($T$) at each interface is obtained by multiplying the potential transmittance $\psi$ of each layer sequentially from the top layer, and $\psi_j$ is given by

$$\psi_j = \frac{\mathrm{Re}(Y_j)}{\mathrm{Re}(Y_{j-1})|\cos\delta_j + iY_j \sin\delta_j / N_j|^2}. \qquad (4)$$

In Fig. 2, $T$ at each interface is indicated and, from $R$ and $\psi$ of each layer, the absorptance of the $j$th layer ($A_j$) is determined as follows:[15,17]

$$A_j = (1-R)(1-\psi_j)\prod_{g=1}^{j-1}\psi_g. \qquad (5)$$

For optically thick c-Si wafers (~150 μm), however, the above calculation procedure cannot be employed because the optical response in this case needs to be calculated under the incoherent condition. More specifically, for optically thick incoherent layers, the optical interference is not observed as the phase information is lost completely by the time-varying phase of light traveling a long distance.[30] Quite fortunately, the calculation procedure for a coherent/incoherent multilayer model has already been established within the Fresnel approach (or transfer matrix method).[31,32] In this method, the phase $\delta$ expressed by Eq. (2) is changed intentionally so that the optical interference effect is eliminated by averaging out the coherent optical response.

By following this approach, we have developed a continuous phase approximation (CPA) method in which incoherent light absorption is described within the above optical admittance calculation by varying the phase of the incoherent layer continuously. In the CPA method, $\delta$ of the incoherent layer is expressed by

$$\delta_p = \frac{2\pi N_j d_j}{\lambda} + \frac{p}{m}\pi, \qquad (6)$$

where $m$ is a total number of the assumed waves and $p$ is the sequential number of the individual wave ($p = 0, 1...m-1$). In Eq. (6), the term $p\pi/m$ indicates a phase added intentionally to $\delta$ of Eq. (2). When the phase is modified, the peak and valley positions of the propagating waves change [see $\psi_j(\delta_p)$ in Fig. 2]. Thus, if all the waves having slightly different $\delta$ values are integrated, the optical interference fringes disappear and the incoherent optical response can be reproduced, as reported previously.[31,32]

As a result, by modifying Eq. (5), the absorptance of the incoherent layer ($A_{j,\mathrm{inc}}$) is described by

$$A_{j,\mathrm{inc}} = \frac{1}{m}\sum_{p=0}^{m-1}(1-R)[1-\psi_j(\delta_p)]\prod_{g=1}^{j-1}\psi_g \qquad (7)$$

$$= (1-R)(1-\psi_{\mathrm{eff}})\prod_{g=1}^{j-1}\psi_g,$$

where $\psi_{\mathrm{eff}}$ indicates the effective $\psi$ of the incoherent layer:

$$\psi_{\mathrm{eff}} = \frac{1}{m}\sum_{p=0}^{m-1}\psi_j(\delta_p). \qquad (8)$$

The above equation confirms that $\psi_{\mathrm{eff}}$ is a simple average of different $\psi_j(\delta_p)$ values. The $R$ of the above incoherent model [i.e., $R$ in Eq. (7)] is further expressed using Eq. (3):

$$R_{\mathrm{CPA}} = \frac{1}{m}\sum_{p=0}^{m-1}\left|\frac{1-Y_0(\delta_p)}{1+Y_0(\delta_p)}\right|^2, \qquad (9)$$

where $R_{\mathrm{CPA}}$ shows the reflectance obtained from the CPA method. By applying the above procedure, the light absorption in a complex multilayer structure with coherent and incoherent layers can be calculated rather easily. If 100% carrier collection is further assumed for the incoherent absorber layer, the corresponding EQE spectrum is obtained from Eq. (7) as $A_{j,\mathrm{inc}}(\lambda)=EQE(\lambda)$. In the actual EQE calculation, the c-Si absorber in the optical model is divided into c-Si sublayers with a thickness of less than 10 μm, as otherwise the computer calculation becomes quite difficult due to quite large $\delta_p$ values.

## B. Calculation of textured solar cells

As confirmed previously,[33-36] thin film structures formed on the {111} facets of c-Si pyramid textures can be modeled using a simple coherent optical model. In fact, when a $SiN_x$ or an ITO layer (~70 nm) is deposited on the pyramid-shaped random texture, the surface shows a blue color, which corresponds to the interference color of the thin film structure. In other words, light scattering within the thin layers is rather small and the overall near-surface optical response is described by the coherent condition. In addition, when the specular light-reflection component of the textures is measured by spectroscopic ellipsometry (SE) using a tilt-angle optical configuration, a-Si:H/c-Si solar cell structures deduced by SE show excellent agreement with those determined by transmission electron microscopy.[34,35]

Figure 3(a) schematically shows the light transmission in an ITO/a-Si:H/c-Si front texture. From experiments, the top angle of the pyramid-shaped texture is confirmed to be 80° (Refs. 34 and 35). The $n$ values of ITO, a-Si:H and c-Si in the region just above the band gap ($E_g$) of c-Si ($\lambda$=1100 nm) are $n$=1.7 (Ref. 25), $n$=3.6 (Ref. 37), and $n$=3.5 (Ref. 38), which result in transmission angles of 27°, 12° and 12°, respectively [see Fig.

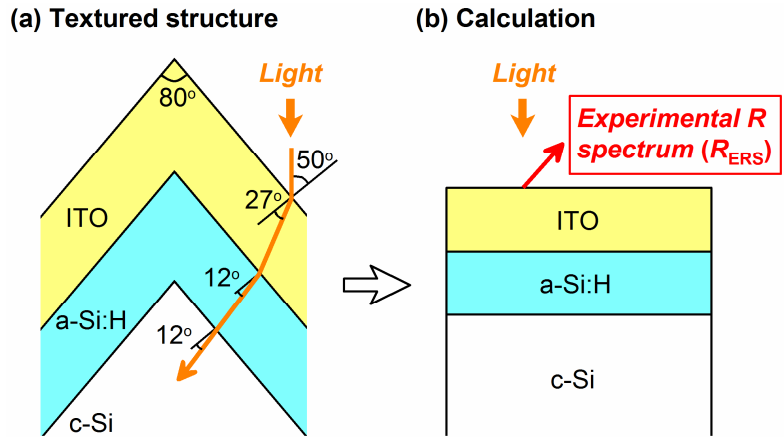

FIG. 3. (a) Optical transmission in an ITO/a-Si:H thin film structure formed on the {111} facet of a pyramid-shaped c-Si texture and (b) calculation method of the optical absorption in the textured structure. In this method, a flat optical model is applied while using the experimental reflectance spectrum ($R_{ERS}$) obtained from an actual textured solar cell.

3(a)]. The transmission angles within the thin layers are close to the normal to the {111} texture-facet plane. Accordingly, based on the above results, we approximate the absorptance of the TCO and a-Si:H layers incorporated into the textured solar cells by assuming the normal incidence within the simple coherent condition [see Fig. 3(b)]. Quite fortunately, for this calculation, the identical calculation procedure described in the previous section (Sec. III A) can be employed. Nevertheless, the c-Si textures reduce $R$ notably in the visible region, if compared with flat c-Si structures,[11] and this antireflection effect should be taken into account. In this study, to estimate the optical absorption in c-Si textured solar cells accurately, the reflectance spectra obtained experimentally from the actual solar cells are adopted for the EQE analyses. These analyses can be implemented quite easily by replacing $R$ in Eq. (7) with $R_{ERS}$, where $R_{ERS}$ represents $R$ of the experimental reflectance spectra. In other words, if the internal quantum efficiency ($IQE$) obtained using a flat optical model is $IQE_{flat}$, the EQE of the textured structure ($EQE_{tex}$) is approximated as

$$EQE_{tex} = (1 - R_{ERS})IQE_{flat}. \tag{10}$$

The above procedure simplifies the EQE analysis of textured c-Si solar cells drastically. In the EQE analysis of the PERL solar cell, on the other hand, the rear interface structure is assumed to be uniform by neglecting the optical contribution of the local $n^+$ region, as the area fraction of the $n^+$ region is rather small (5%).[22]

**C. Optical constants of solar cell materials**

For all the optical analyses in this study, the optical constants of c-Si reported by Herzinger et al.[38] were adopted, whereas the optical data of the Al and Ag rear electrodes were taken from Ref. 39 and Ref. 40, respectively. In the analyses of the a-SiO:H/c-Si solar cells [Figs. 1(a) and 1(b)], the optical constants of ITO (Ref. 25) and a-SiO:H (Ref. 41), extracted from samples fabricated using similar growth conditions, were used. In particular, the optical carrier concentration ($N_{opt}$) of the front and rear ITO layers in the devices is $5 \times 10^{20}$ cm$^{-3}$ (Refs. 13 and 25).

For the textured a-Si:H/c-Si solar cell in Fig. 1(c), the optical constants of the a-Si:H and ITO layers incorporated into the solar cells have been characterized by SE[9] and these optical data were adopted for the calculation. In the analysis of Fig. 1(c), however, we assumed that the optical properties of the a-Si:H i-n layers are identical. Moreover, based on Ref. 10, the carrier concentrations of the front and rear ITO layers are further assumed to be $2.4 \times 10^{20}$ cm$^{-3}$ and $1.7 \times 10^{19}$ cm$^{-3}$, respectively.

For the EQE analysis of the dopant-free solar cell [Fig. 1(d)], the reported optical constants of a-Si:H (Ref. 9), ITO ($N_{opt}$=2.4×10$^{20}$ cm$^{-3}$ of Ref. 10), IOH ($N_{opt}$=2.1×10$^{20}$ cm$^{-3}$ of Ref. 42) and LiF (Ref. 43) were used, whereas we employed a MoO$_x$ dielectric function extracted from a sputtered MoO$_x$ layer[40] for the calculation.

In the EQE analysis of the PERL cell [Fig. 1(e)], the dielectric functions of a wide-gap Si$_3$N$_4$ layer[44] and an Al$_2$O$_3$ layer prepared by atomic layer deposition[45] were employed. The parameterization of the Si$_3$N$_4$ dielectric function using the Tauc-Lorentz model[46] leads to $A$=150.733 eV, $E_0$=8.416 eV, $C$=3.962 eV, $E_g$=4.825 eV and $\varepsilon_1(\infty)$=1.478. For the actual calculation of the Si$_3$N$_4$ dielectric function, these parameters were employed. For the SiO$_2$ passivation layer, we used the optical constants of a thermal oxide formed on c-Si.[38]

## IV. RESULTS

### A. Analysis of flat heterojunction solar cells

To confirm the validity of the CPA method, the EQE spectrum of the DH a-SiO:H/c-Si solar cell without texture [i.e., Fig. 1(b)] was analyzed first. Figure 4 shows the experimental EQE spectrum of this solar cell reported in Ref. 14 (open circles) and the calculated EQE spectra (solid lines). In this analysis, the EQE of the c-Si was deduced from Eq. (7) using $m$=13 assuming 100% carrier collection ($A_{j,inc}$=EQE). The red line represents the EQE spectrum calculated from the CPA method, whereas the EQE spectra obtained using selected $\delta_p$ values (i.e., $p$=0, 1, 2, 12) are also shown (see also the enlarged figure). For the choice of $m$, we find that (i) a prime number and (ii) a larger $m$ value are favorable to eliminate the interference fringes effectively.
As shown in Fig. 4, when the EQE is calculated using a fixed $\delta_p$, quite sharp optical interference appears particularly in a low light absorption region of c-Si ($\lambda$>1000 nm) and the interference pattern changes systematically with $\delta_p$. If the average optical absorption is calculated from Eq. (7), therefore, all the sharp absorption features are averaged out and a quite smooth incoherent spectrum can be obtained.

Figure 5 shows the EQE analysis result for the flat DH a-SiO:H/c-Si solar cell. The experimental EQE (open circles) and the EQE calculated from the CPA method (red line) are consistent with Fig. 4, whereas the black lines indicate the CPA-derived reflectance spectrum ($R_{CPA}$) and absorptance spectra of the solar-cell component layers. In Fig. 5, the calculated EQE spectrum shows almost perfect agreement with the experimental EQE spectrum in the wide $\lambda$ region. As a result, $J_{sc}$ obtained from the CPA

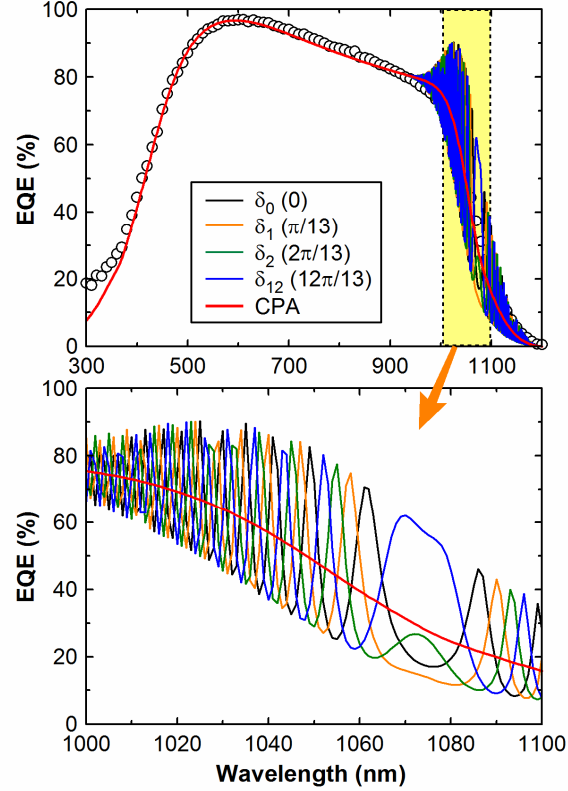

FIG. 4. Experimental EQE spectrum (open circles) and calculated EQE spectra (solid lines) of the flat a-SiO:H/c-Si solar cell (DH) shown in Fig. 1(b). The experimental spectrum was taken from Ref. 14. For the EQE spectra calculated using fixed $\delta_p$ values, only the results of $p$=0, 1, 2 and 12 ($m$=13) are shown for clarity. The values inside the parentheses show $\delta$ added intentionally [see Eq. (6)]. The red line represents the EQE spectrum calculated by applying the CPA method. The enlarged spectra in the range of $1000 \leq \lambda \leq 1100$ nm are also shown.

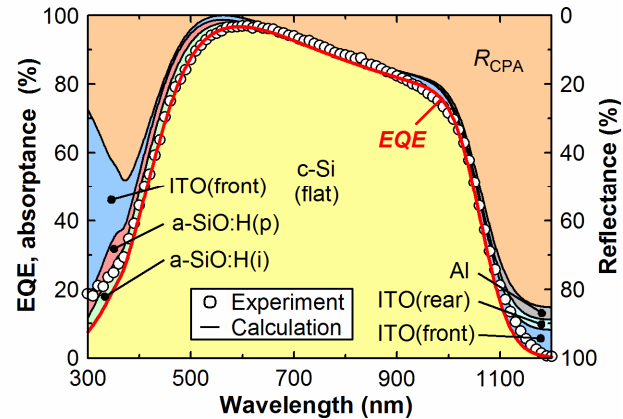

FIG. 5. EQE analysis result for the flat a-SiO:H/c-Si solar cell (DH) shown in Fig. 1(b). The experimental EQE (open circles) and the calculated EQE (red line) are consistent with Fig. 4. The black lines indicate the reflectance spectrum and absorptance spectra of the solar-cell component layers, deduced from the CPA method.

calculation (34.9 mA/cm$^2$) agrees quite well with the experimental $J_{sc}$ (35.4 mA/cm$^2$).

We emphasize that the above EQE analysis was implemented without using any fitting parameters. In particular, the a-SiO:H/c-Si solar cell was fabricated by real-time control of the a-SiO:H layer thicknesses using SE with an accuracy of ~1 Å,[47] and the structural uncertainty is quite small for this solar cell. The excellent agreement observed between the experimental and calculated EQE spectra confirms that the a-SiO:H layers are "dead layers" that allow almost no carrier extraction.

The above EQE analysis shows clearly that the parasitic absorption in the front structure [i.e., ITO/a-SiO:H(p-i)] is relatively large with a total loss of 2.3 mA/cm$^2$, while the rear structure [i.e., a-SiO:H(i-n)/ITO/Al] shows a very small optical loss of 0.4 mA/cm$^2$. In this solar cell, however, the largest $J_{sc}$ loss occurs by the reflectance loss (8.9 mA/cm$^2$) due to the flat device structure.

To justify the CPA approach further, flat a-Si:H/c-Si solar cells reported in Ref. 11 were analyzed. These solar cells have a structure of ITO/a-Si:H p (5 nm)/a-Si:H i (5 nm)/c-Si (280 μm)/a-Si:H n (9 nm)/Ag, and a series of the solar cells were made by varying the ITO layer thickness (53~93 nm). For the EQE analyses, dielectric functions of a-Si:H processed at 130 °C (Ref. 37) and ITO ($N_{opt}$=4.9×10$^{20}$ cm$^{-3}$) of Ref. 25 were employed. Figure 6 summarizes the experimental EQE and reflectance spectra (open symbols) and the corresponding spectra calculated based on the CPA method (solid lines). For all the EQE and reflectance spectra, remarkable agreement has been observed. Thus, $R_{CPA}$ obtained from Eq. (9) provides good matching to the experimental result. The above result further supports that only the photocarriers generated within c-Si contribute to $J_{sc}$ and those created within the a-Si:H layers are lost by recombination. Our result is slightly different from those of earlier studies in which slight carrier extraction (~30%) from a-Si:H i layers is reported to occur.[9,11]

**B. Modeling of carrier recombination**

By extending the CPA method, we have further characterized the carrier recombination observed in the SH a-SiO:H/c-Si solar cell without a BSF structure[14] [i.e., Fig. 1(a)]. Figure 7 compares the EQE spectra obtained from the SH and DH a-SiO:H/c-Si solar cells of Figs. 1(a) and 1(b). In Fig. 7, the experimental EQE spectrum (open circles) of the SH solar cell shows the lower EQE response in the long $\lambda$ region ($\lambda$ > 700 nm), compared with the DH solar cell, due to the effect of the rear interface recombination. In this case, therefore, the recombination effect needs to be incorporated into the analysis.

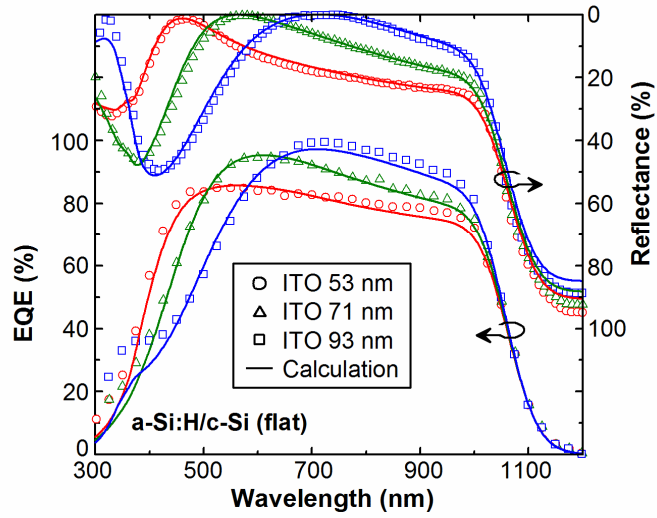

FIG. 6. EQE analysis results obtained from the flat a-Si:H/c-Si solar cells (DH) with different ITO layer thicknesses. The experimental EQE and reflectance spectra reported in Ref. 11 (open symbols) and the corresponding spectra calculated based on the CPA method (solid lines) are shown.

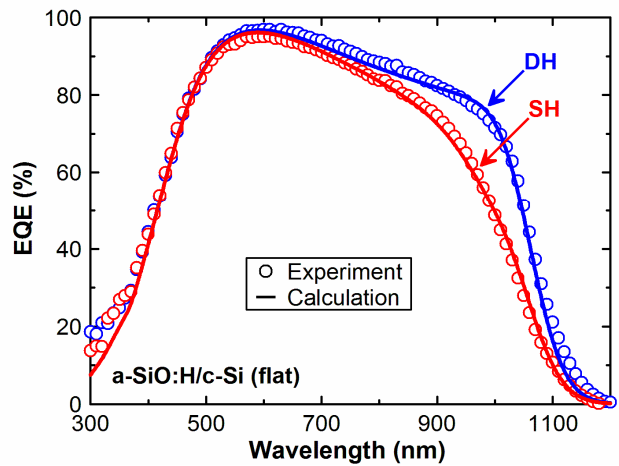

FIG. 7. EQE spectra of the SH and DH a-SiO:H/c-Si solar cells fabricated using flat c-Si substrates [see Figs. 1(a) and 1(b)]. The experimental EQE spectra reported in Ref. 14 (open circles) and the calculated EQE (solid lines) are shown. The result of the DH solar cell is consistent with Fig. 5. For the EQE analysis of the SH solar cell, the carrier recombination at the rear interface has been taken into account.

We have modeled the rear interface recombination by considering the carrier collection efficiency $H(x)$:

$$H(x) = \frac{K \exp\left(\dfrac{d-x}{L_p}\right) + \exp\left(-\dfrac{d-x}{L_p}\right)}{K \exp\left(\dfrac{d}{L_p}\right) + \exp\left(-\dfrac{d}{L_p}\right)}. \qquad (11)$$

The above equation has been derived assuming an ideal p-n junction solar cell using the inverse Laplace transformation.[48] In Eq. (11), $x$ shows the depth from the front interface of the c-Si, whereas $d$ and $L_p$ indicate the c-Si wafer thickness and the diffusion length of the p-type minority carrier, respectively. The $K$ in Eq. (11) is a coefficient given by

$$K = \frac{1 + S_p L_p / D_p}{1 - S_p L_p / D_p} \qquad (12)$$

where $S_p$ and $D_p$ show the surface recombination velocity and diffusion constant of holes, respectively. If this equation is applied, the carrier collection at the depth $x$ can be determined using $S_p$ and $L_p$ as variables. For the calculation, we adopted $D_p = 12.95$ cm$^2$/s assuming hole mobility of $\mu_p = 500$ cm$^2$/(Vs) at a carrier concentration of $1 \times 10^{16}$ cm$^{-3}$ (Ref. 49). In our analysis, the effect of the depletion layer was neglected since (i) the depletion layer thickness is much thinner than $d$ and (ii) its effect on the EQE is minor. The electric-field-assisted carrier collection in the depletion layer can be modeled rather easily assuming $H=1$ in this region.[48]

Figure 8 shows the variations of $H(x)$ with (a) $S_p$ and (b) $L_p$, obtained from Eq. (11). In the figures, the position of $x=0$ indicates the a-SiO:H(i)/c-Si interface, and $H(x)$ was calculated using $d=300$ μm. In Fig. 8(a), $S_p$ is varied while fixing $L_p$ to 1000 μm, whereas $L_p$ is varied with a fixed $S_p$ of $10^3$ cm/s in Fig. 8(b). If $L_p=\infty$ and $S_p=0$, $H(x)$ shows a constant value of 100% [dotted line in Fig. 8(a)]. With increasing $S_p$, however, $H$ at the rear interface (i.e., $x=300$ μm) decreases and becomes zero at $S_p \geq 10^5$ cm/s. When $L_p$ is varied, $H(x)$ in the c-Si bulk region decreases significantly due to the limited carrier collection. Accordingly, $S_p$ and $L_p$ can be estimated separately if $H(x)$ is determined.

In our recombination analysis, $H(x)$ is incorporated directly into the EQE analysis according to

$$EQE(\lambda) = \int A_{inc}(x,\lambda) H(x) dx, \qquad (13)$$

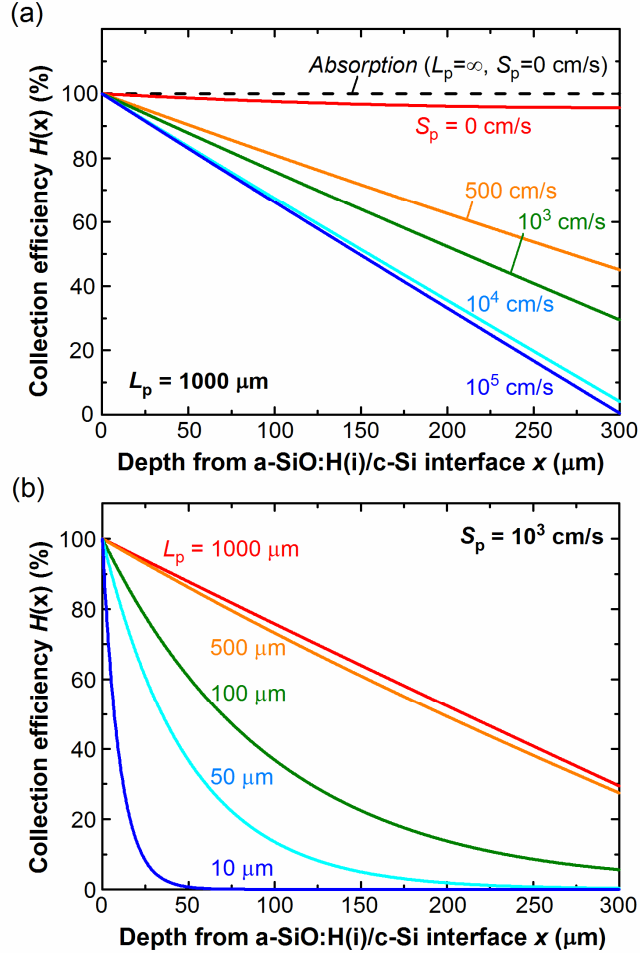

FIG. 8. Variations of $H(x)$ with (a) $S_p$ and (b) $L_p$, obtained from Eq. (11). In (a), a fixed $L_p$ value of 1000 μm is assumed, whereas $S_p$ is fixed at $10^3$ cm/s in (b). The dotted line shows the case of 100% carrier collection ($L_p=\infty$ and $S_p=0$ cm/s).

where $A_{\text{inc}}(x, \lambda)$ is the absorptance of the incoherent c-Si absorber at the depth $x$ and $\lambda$. The $A_{\text{inc}}(x, \lambda)$ can be calculated rather easily by dividing the c-Si layer into many sublayers having the same optical constants in the optical model. In the actual analysis of the SH solar cell, the 300-μm-thick c-Si absorber was divided into a total of 1500 sublayers with a thickness of 200 nm.

By applying Eq. (13), we extracted the parameters ($S_p$, $L_p$) from the EQE fitting analysis. The red line in Fig. 7 represents the result obtained from this fitting analysis and the calculated EQE spectrum shows the excellent agreement with the experimental spectrum when $L_p$=1000 μm and $S_p$=$10^3$ cm/s. However, $H(x)$ of this solar cell is essentially limited by $S_p$ due to the lack of the BSF structure and the EQE spectrum shows little change with $L_p \geq$1000 μm. The above result shows clearly that detailed analysis of the carrier recombination can be performed by combining the depth-resolved carrier recombination analysis with the CPA method.

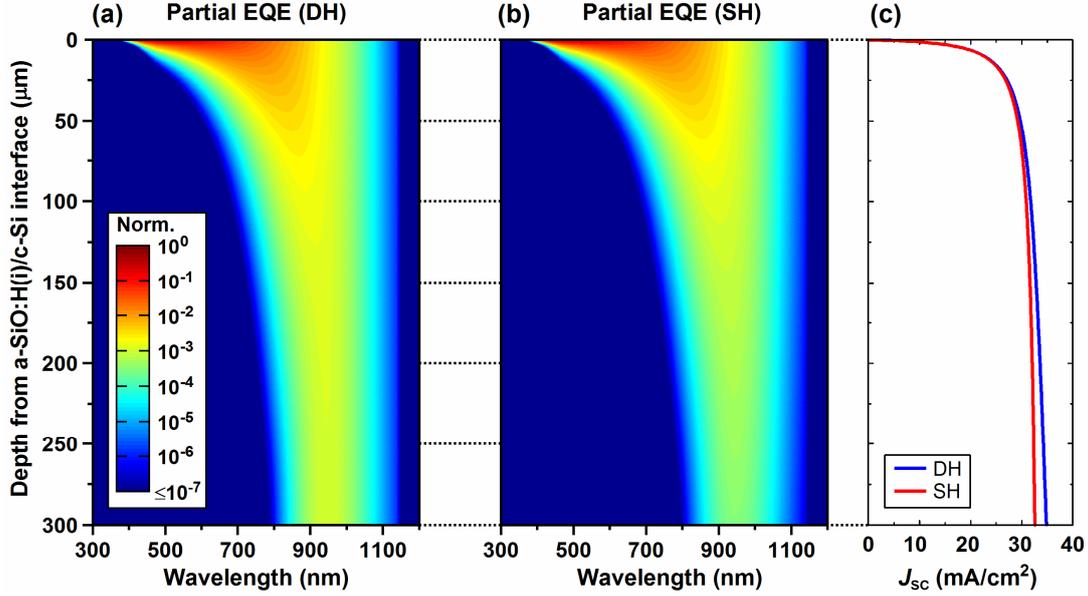

FIG. 9. Normalized partial EQE obtained at different depths from the a-SiO:H(i)/c-Si interface and wavelengths in (a) the DH and (b) the SH a-SiO:H/c-Si solar cells and (c) the integrated $J_{sc}$ for the depth from the a-SiO:H/c-Si interface. These partial EQE values correspond to the EQE spectra shown as the solid lines in Fig. 7.

Figure 9 shows the normalized partial EQE of (a) the DH and (b) the SH a-SiO:H/c-Si solar cells and (c) the integrated $J_{sc}$ for the depth from the a-SiO:H(i)/c-Si interface (i.e., $x$). The partial EQE represents an EQE value obtained at specific ($x$, $\lambda$) values. If the partial EQE spectra obtained at different depths are integrated, the EQE spectra indicated by the solid lines in Fig. 7 are obtained. In Figs. 9(a) and 9(b), the partial EQE is indicated using logarithmic scale. The calculation result reveals that, in the region of $\lambda$ < 700 nm, the partial EQE decreases rapidly up to $x$=50 μm due to the strong light absorption in c-Si. In contrast, at $\lambda$ > 800 nm, the weak indirect absorption in this region leads to the quite uniform carrier generation throughout the entire absorber.

Thus, the influence of the rear-interface carrier recombination appears predominantly in this $\lambda$ region. As shown in Fig. 8(a), $H$ decreases almost linearly with $x$ and the partial EQE of the SH solar cell decreases in the longer $\lambda$ region, compared with the DH solar cell. In particular, since $L_p$ (1000 μm) is larger than the absorber thickness (300 μm), the intense rear-interface recombination reduces the EQE response of the SH solar cell notably.

In Fig. 9(c), the integrated $J_{sc}$ values relative to $x$ are shown. The integrated $J_{sc}$ values of the SH and DH solar cells are almost identical up to $x \sim 50$ μm but the recombination in the SH solar cell hinders the increase in $J_{sc}$ at $x > 50$ μm, resulting in the $J_{sc}$ reduction of 2.5 mA/cm$^2$. The result of Fig. 9 also shows that a stronger optical confinement is critical to achieve high efficiencies when a thinner c-Si wafer is used.

## C. Analysis of textured solar cells

To validate our EQE analysis procedure established for textured c-Si solar cells, a series of standard a-Si:H/c-Si heterojunction solar cells with different a-Si:H layer thicknesses have been characterized. Figure 10 shows the EQE spectra of the textured a-Si:H/c-Si solar cells fabricated by varying the a-Si:H p layer thickness (a) without the a-Si:H i layer and (b) with the a-Si:H i layer. In this figure, the symbols show reported experimental results[9] and the solid lines indicate our calculation results. The basic structure of the solar cells is identical to that of Fig. 1(c). Since only the EQE and $R$ spectra in the short $\lambda$ region ($\lambda \leq 600$ nm) were reported in Ref. 9, for these EQE analyses, we employed a simplified optical model consisting of ITO(70 nm)/a-Si:H(p)/[a-Si:H(i)]/c-Si without considering the rear interface structure. In particular, all the light in the region of $\lambda \leq 600$ nm is absorbed completely within the c-Si (see Fig. 9) and the effect of the rear structure can be neglected. In the analysis of Fig. 10(b), the i layer thickness is assumed to be constant (5 nm).

In Fig. 10, the calculated EQE spectra show remarkable agreement with those confirmed experimentally and the reduction of the short-$\lambda$ EQE with increasing the p layer thickness is reproduced quite well. It should be emphasized that there are essentially no adjustable analytical parameters in these analyses and the EQE spectra are calculated simply from the layer thicknesses and the optical constants of the layers. Furthermore, our calculation results are quite consistent with those obtained from the ray-tracing analyses performed for the same solar cells.[9] Accordingly, although the experimental $R$ spectrum is always necessary in our approach, our technique provides a quite quantitative estimation of the optical loss in the textured c-Si solar cells.

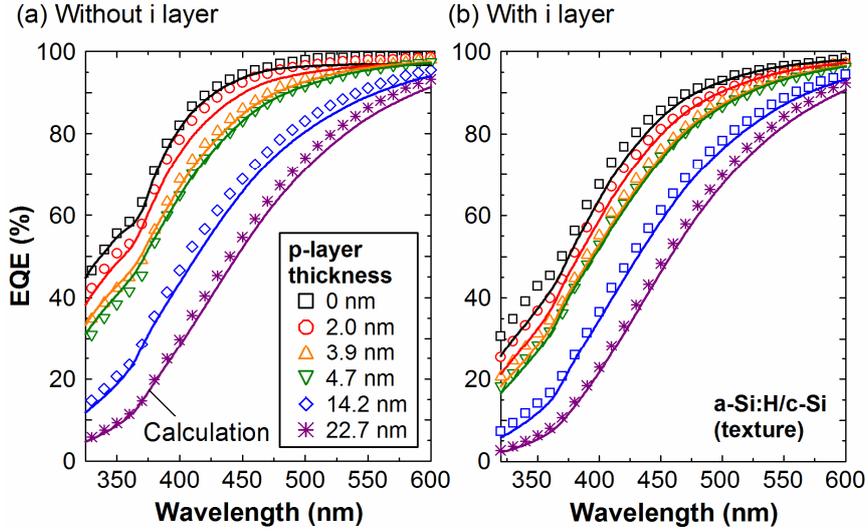

FIG. 10. EQE analysis results obtained for the textured a-Si:H/c-Si solar cells fabricated by varying the a-Si:H p layer thickness (a) without the a-Si:H i layer and (b) with the a-Si:H i layer. The experimental EQE spectra reported in Ref. 9 (symbols) and the calculated EQE (solid lines) are shown.

Conversely, from the EQE analysis of this region, the a-Si:H layer thickness can be estimated assuming no carrier extraction from the a-Si:H layers.

Figure 11 summarizes the complete EQE analyses performed for the textured solar cells shown in Fig. 1: EQE analyses performed for (a) the standard a-Si:H/c-Si, (b) dopant-free MoO$_x$/c-Si and (c) PERL solar cells, and (d) experimental EQE and $R$ spectra of the solar cells. The open circles and squares show the EQE and $R$ spectra obtained experimentally,[20-22] whereas the solid lines indicate the calculated EQE spectra (red lines) and the absorptance spectra of each component layer (black lines). Rather surprisingly, the calculated EQE spectra shown in Figs. 11(a)-(c) indicate almost perfect fitting in the whole analyzed region particularly for the a-Si:H/c-Si and PERL solar cells, even though quite simple EQE analyses were performed using flat optical models. In the analyses of the a-Si:H/c-Si and MoO$_x$/c-Si solar cells, however, the a-Si:H layer thicknesses were slightly reduced, compared with the reported thicknesses [see Figs. 1(c) and 1(d)], since otherwise the calculated EQE becomes notably lower than the experimental EQE in the short λ region of 300≤λ≤600 nm. The remarkable agreement observed between the experimental and calculated EQE spectra indicates that the carrier recombination at the rear interface is negligible in these solar cells due to the presence of the BSF structures.

The result of Fig. 11(d) confirms that the EQE in the short $\lambda$ region is limited in the a-Si:H/c-Si and MoO$_x$/c-Si solar cells because of the parasitic light absorption in the TCO and a-Si:H layers. In the longer $\lambda$ region, on the other hand, the EQE response of all the solar cells is quite similar. The $R$ of the PERL cell is, however, quite large, compared with the heterojunction solar cells, most likely due to the flat rear interface structure [see Fig. 1(e)].

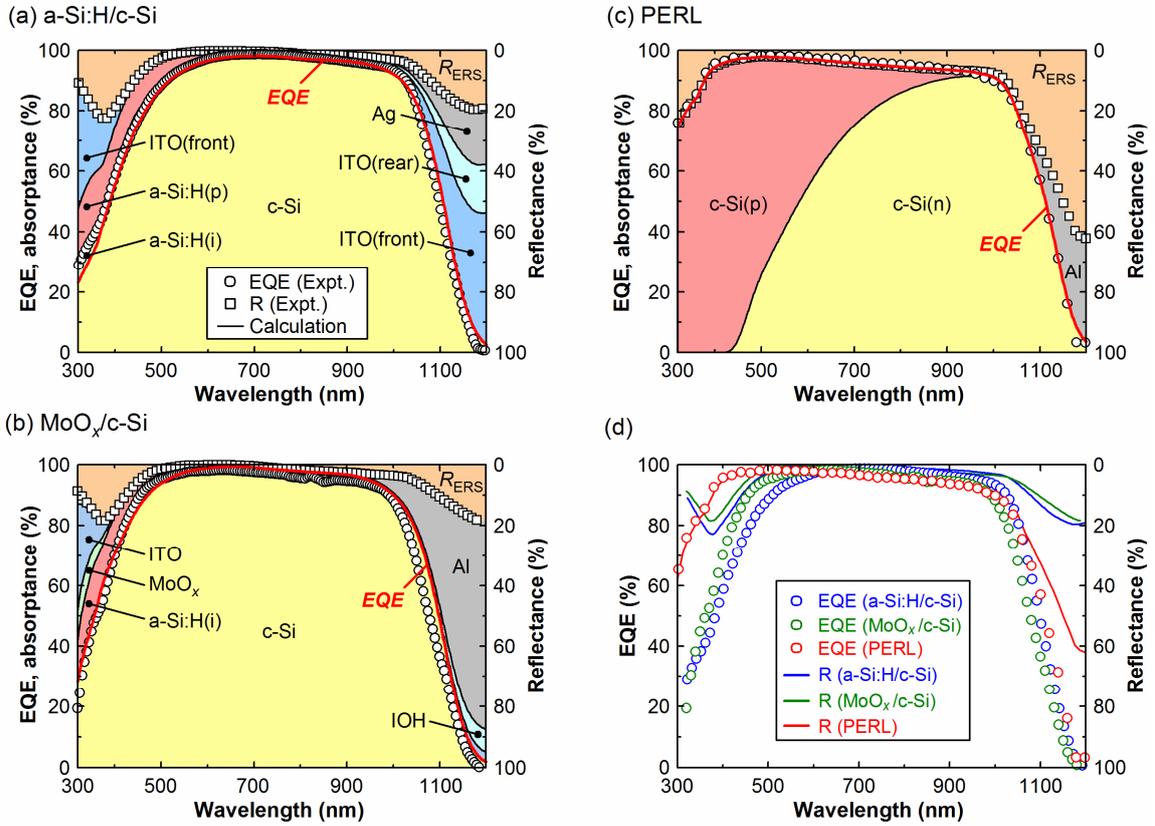

FIG. 11. EQE analysis results obtained from the textured c-Si solar cells shown in Fig. 1: (a) standard a-Si:H/c-Si, (b) dopant-free MoO$_x$/c-Si and (c) PERL solar cells, and (d) comparison of the experimental EQE and $R$ spectra. The open circles and squares show the experimental EQE and $R$ spectra,[20-22] respectively. In (a)-(c), the red line indicates the calculated EQE spectrum, whereas the black lines show the absorptance spectra of each component layer.

## V. DISCUSSION

**A. Carrier loss mechanisms in textured c-Si solar cells**

Based on the EQE analyses of Fig. 11, the optical losses in each solar cell have been determined. Figure 12 summarizes the $J_{sc}$ losses (gains) in the solar-cell component (absorber) layers in the textured (a) a-Si:H/c-Si, (b) MoO$_x$/c-Si and (c) PERL solar cells. In the figures, the numerical values represent the current densities in units of mA/cm$^2$ and these values were estimated from the calculated spectra (solid lines in Fig. 11). The maximum $J_{sc}$ attainable under AM1.5G condition in the region of $300 \leq \lambda \leq 1200$ nm is 46.5 mA/cm$^2$ and the PERL solar cell shows the highest optical gain of 88.8%, while the a-Si:H/c-Si solar cell shows a lower gain of 85.6%.

In the a-Si:H/c-Si solar cell, a relatively large absorption loss occurs in the front structure (3.5 mA/cm$^2$), whereas the $J_{sc}$ loss in the rear structure is rather small (1.3 mA/cm$^2$). In the dopant-free solar cell, the a-Si:H p layer is removed but the parasitic absorption still occurs in the a-Si:H i layer. In particular, the $J_{sc}$ loss generated by this a-Si:H i layer (0.8 mA/cm$^2$) is comparable to that induced by the a-Si:H p-i layers in the a-Si:H/c-Si (1.4 mA/cm$^2$). Thus, the improvement of the short-$\lambda$ EQE response in the dopant-free solar cell is rather limited.

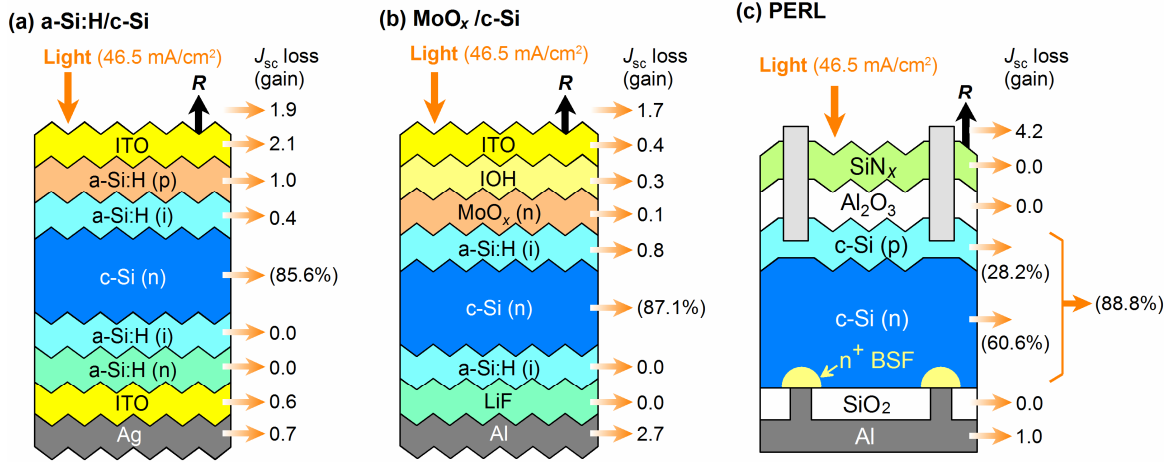

FIG. 12. $J_{sc}$ losses (gains) in the solar-cell component (absorber) layers in the textured (a) a-Si:H/c-Si, (b) MoO$_x$/c-Si and (c) PERL solar cells. The numerical values represent the corresponding current densities in units of mA/cm$^2$. The maximum $J_{sc}$ attainable under AM1.5G condition is 46.5 mA/cm$^2$ ($\lambda$=300-1200 nm) from which the optical gain is calculated as the ratio of output $J_{sc}$ to the maximum $J_{sc}$.

Moreover, in the a-Si:H/c-Si solar cell, the optical losses in the rear a-Si:H i-n layers are zero. This effect can be interpreted by high $E_g$ of a-Si:H layers (~1.7 eV), which lead to strong light absorption only in the region of $\lambda$<730 nm. As confirmed from Fig. 9, all the light in this region is absorbed in the c-Si upper layer and thus the parasitic absorption in the rear a-Si:H layers is negligible.

In the $MoO_x$/c-Si solar cell that incorporates the high-mobility IOH layer, the optical loss induced by the front TCO is well suppressed, compared with the a-Si:H/c-Si solar cell. In the dopant-free solar cell, however, quite strong parasitic absorption occurs in the Al rear electrode [see also Fig. 11(b)]. This shows an important fact for the light absorption in solar cells; i.e., the absorptance of the absorber layer in a multilayer solar cell is essentially governed by the relative magnitude of the absorption coefficient ($\alpha$) and thickness of the component layers. In other words, even when the front optical loss is reduced, the light absorption in the indirect-transition c-Si absorber may not increase significantly, if another component layer has a higher $\alpha$ value than that of c-Si. In the case of the a-Si:H/c-Si solar cell, for example, the contributions of the parasitic absorption observed at $\lambda$ > 1000 nm are roughly equal for the front ITO, rear ITO and Ag, although the light absorption in the front layer tends be larger than the rear layer. However, if the front parasitic absorption is removed [i.e., Fig. 12(b)], the light absorption in the rear metal electrode becomes dominant as $\alpha$ of metals is far larger than that of c-Si. Accordingly, to improve the longer $\lambda$ response in c-Si solar cells, enhanced light scattering in the front texture or the increase in the c-Si thickness is necessary.

In the PERL solar cell, the optical loss in the front layers is eliminated completely and the EQE is equal to $1-R$ at $\lambda \leq 1000$ nm. Furthermore, there is no carrier loss in the p-type emitter and its contribution to $J_{sc}$ (13.1 mA/cm$^2$) accounts for 28% of the total $J_{sc}$. In this solar cell, the largest optical loss occurs by the light reflection (4.2 mA/cm$^2$), which is notably larger than those in the heterojunction solar cell (~2 mA/cm$^2$) due to the flat rear interface structure. A ray-tracing simulation of c-Si solar cells has already confirmed the clear increase in $J_{sc}$ by double-sided texturing,[4] but the rear texturing makes the formation of the PERL structure more difficult. As mentioned above, although the optical gain of the PERL cell is high, one disadvantage of this solar cell is a lower $V_{oc}$, compared with the heterojunction solar cells.

## B. Effect of c-Si thickness

In the above EQE analyses, the c-Si wafer thicknesses of the actual solar cells were used. However, if the transmission angle depicted in Fig. 3 is assumed, the effective optical pass length within the textured c-Si is expected to increase by 27% [i.e., $1/\cos(50°-12°)$], although this effect is neglected completely in our analysis. To find the effect of c-Si thickness on EQE, we have simulated the EQE spectra of textured a-Si:H/c-Si solar cells having different wafer thicknesses. For the simulations, we assumed the a-Si:H/c-Si structure of Fig. 1(c). Unfortunately, the complete device simulation of c-Si textured structures is difficult in our approach due to the necessity of the corresponding $R$ spectra, and we employed a fixed $R$ spectrum for all the calculations. Since $R$ varies with the c-Si thickness, the optical simulation performed here is hypothetical. However, the change in $R$ observed in a-Si:H/c-Si solar cells in a thickness range of 100-250 μm is rather small ($\Delta R<5\%$)[10] and the effect of $R$ is expected to be minor.

Figure 13 shows the variation of EQE spectrum with c-Si thickness obtained from the optical simulation. The solid lines show the simulation result, whereas the open symbols indicate the experimental data of Fig. 11(a).[20] With increasing the wafer thickness, the EQE in the longer $\lambda$ increases gradually and the experimental EQE shows good agreement when the c-Si thicknesses in the simulations are 150~200 μm. These thicknesses are slightly thinner than the actual c-Si thickness (230 μm). If the enhanced optical pass length by the inclined transition angle (i.e., 27%) is considered for the 230-μm-thick substrate, the effective c-Si thickness becomes ~290 μm. In this case, the experimental EQE becomes slightly smaller than the simulation result, suggesting the slight recombination at the rear interface. Nevertheless, the above EQE simulation could

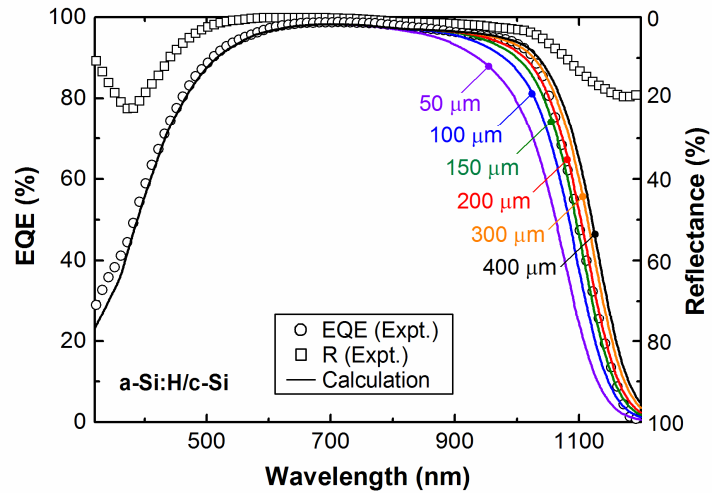

FIG. 13. Variation of EQE spectrum with the c-Si thickness obtained from the optical simulation. The open circles and squares show the experimental EQE and $R$ spectra reported in Ref. 20, respectively. In the EQE simulations, $R$ obtained from the

a-Si:H/c-Si solar cell with a 230-μm-thick c-Si substrate[20] is assumed to be constant.
be too simple to discuss very small carrier losses observed in the longer $\lambda$ region. Thus, the effect of the c-Si substrate thickness in textured solar cells needs to be clarified further based on the EQE analysis results obtained with the variation of the c-Si thickness.

**VI. Conclusion**

We established a global EQE analysis method that can be applied for quantitative analysis of the optical and recombination losses in various c-Si solar cells. In this calculation scheme, a flat optical model is employed within the framework of the optical admittance method, and the incoherent optical absorption in thick c-Si substrates is expressed by using a procedure reported earlier. We find that the EQE calculation of textured c-Si solar cells can be performed by applying experimental reflectance spectra to the above method. Our approach provides excellent fittings to numerous EQE spectra reported for high-efficiency c-Si solar cells fabricated using flat and pyramid-shaped c-Si substrates. The main advantage of the established method is a very low computational cost and the EQE calculation can be performed quite easily if the optical constants and thicknesses of all the layers are known. Based on the EQE analyses, $J_{sc}$ losses induced by the front light reflection and parasitic light absorption in solar-cell component layers were deduced. Furthermore, an EQE analysis procedure that allows the extraction of the carrier recombination characteristics of the solar cell was developed. Finally, we note that our method can further be applied to determine carrier loss mechanisms in multi-junction solar cells consisting of group III-V compound semiconductors.